\documentclass[aps,prd,twocolumn,nofootinbib,superscriptaddress,preprintnumbers,floatfix]{revtex4-2}
\usepackage{float,dcolumn}
\usepackage{amsmath,amssymb}
\usepackage{mathtools,slashed}
\usepackage[dvipsnames]{xcolor}
\usepackage[caption=false]{subfig}
\usepackage{verbatim}
\usepackage{array}
\usepackage{multirow}
\usepackage{booktabs}
\usepackage{url}
\usepackage{color}
\usepackage[T1]{fontenc}
\usepackage[utf8]{inputenc}
\usepackage{graphicx}
\usepackage{esint}
\usepackage[hypertexnames=false]{hyperref}
\usepackage{atbegshi}
\usepackage{placeins}
\usepackage{tikz}
\usepackage[capitalise, english]{cleveref}
\usepackage{comment}
\usepackage{microtype}

\usepackage{physics}

\usepackage[normalem]{ulem}

\crefname{equation}{Eq.}{Eqs.}
\crefname{section}{Sect.}{Sects.}
\Crefname{section}{Sec.}{Sects.}
\Crefname{appendix}{Appendix}{Appendices}
\newcommand{\rcite}[1]{Ref.~\cite{#1}}
\newcommand{\rrcite}[1]{Refs.~\cite{#1}}

\allowdisplaybreaks

\hypersetup{ 
pdfnewwindow=true,
colorlinks=true,    
linkcolor=blue,
citecolor=blue,
filecolor=blue,
urlcolor=blue
}
\newcommand{\hd}[1]{\par
     \phantomsection\addcontentsline{toc}{section}{#1}%
     {\itshape #1.}---\ignorespaces}

\begin{document}

\preprint{DESY-26-150}
\title{Quark-mass dependence of two-nucleon scattering}

\author{Jorge~Baeza-Ballesteros}
\email{jorge.baeza.ballesteros@desy.de}
\affiliation{John von Neumann-Institut f\"ur Computing NIC, Deutsches Elektronen-Synchrotron DESY, Platanenallee 6, 15738 Zeuthen, Germany}

\author{Andr\'e~Bai\~{a}o-Raposo}
\email{andre.baiaoraposo@ruhr-uni-bochum.de}
\affiliation{Fakult\"at f\"ur Physik und Astronomie, Institut f\"ur Theoretische Physik II, Ruhr-Universit\"at Bochum, 44780 Bochum, Germany}

\author{Davide~Laudicina}
\email{davide.laudicina@ruhr-uni-bochum.de}
\affiliation{Fakult\"at f\"ur Physik und Astronomie, Institut f\"ur Theoretische Physik II, Ruhr-Universit\"at Bochum, 44780 Bochum, Germany}

\author{Malcolm~Lazarow}
\email{mlazarow@berkeley.edu}
\affiliation{Department of Physics, University of California, Berkeley, California 94720, USA}
\affiliation{Nuclear Science Division, Lawrence Berkeley National Laboratory, Berkeley, California 94720, USA}

\author{Thomas R.~Richardson}
\email{thomas.richardson@berkeley.edu}
\affiliation{Department of Physics, University of California, Berkeley, California 94720, USA}
\affiliation{Nuclear Science Division, Lawrence Berkeley National Laboratory, Berkeley, California 94720, USA}

\author{Miguel~Salg}
\email{miguel.salg@unibe.ch}
\affiliation{Institute for Theoretical Physics, Albert Einstein Center for Fundamental Physics, University of Bern, 3012 Bern, Switzerland}

\author{John~Bulava}
\affiliation{Fakult\"at f\"ur Physik und Astronomie, Institut f\"ur Theoretische Physik II, Ruhr-Universit\"at Bochum, 44780 Bochum, Germany}

\author{M.A.~Clark}
\affiliation{NVIDIA Corporation, Santa Clara, California 95051, USA}

\author{Jeremy~R.~Green}
\affiliation{John von Neumann-Institut f\"ur Computing NIC, Deutsches Elektronen-Synchrotron DESY, Platanenallee 6, 15738 Zeuthen, Germany}

\author{Andrew~Hanlon}
\affiliation{Center for Nuclear Research, Department of Physics, Kent State University, Kent, Ohio 44242, USA}

\author{Renwick~J.~Hudspith}
\affiliation{Department of Physics, Carnegie Mellon University, Pittsburgh, Pennsylvania 15213, USA}

\author{Daniel~Mohler}
\affiliation{GSI Helmholtzzentrum f\"ur Schwerionenforschung GmbH, Planckstra{\ss}e 1, 64291 Darmstadt, Germany}
\affiliation{Institut f\"ur Kernphysik, Technische Universit\"at Darmstadt, Schlossgartenstra{\ss}e 2, 64289 Darmstadt, Germany}

\author{Colin~Morningstar}
\affiliation{Department of Physics, Carnegie Mellon University, Pittsburgh, Pennsylvania 15213, USA}

\author{Amy~Nicholson}
\affiliation{Department of Physics and Astronomy, University of North Carolina, Chapel Hill, North Carolina 27516-3255, USA}
\affiliation{Nuclear Science Division, Lawrence Berkeley National Laboratory, Berkeley, California 94720, USA}

\author{Fernando~Romero-L\'opez}
\affiliation{Institute for Theoretical Physics, Albert Einstein Center for Fundamental Physics, University of Bern, 3012 Bern, Switzerland}

\author{Andr\'e~Walker-Loud}
\affiliation{Nuclear Science Division, Lawrence Berkeley National Laboratory, Berkeley, California 94720, USA}
\affiliation{Department of Physics, University of California, Berkeley, California 94720, USA}
\affiliation{NERSC, Lawrence Berkeley National Laboratory, Berkeley, CA 94720, USA}

\author{Hartmut~Wittig}
\affiliation{PRISMA$^{++}$ Cluster of Excellence and Institut f\"ur Kernphysik, Johannes Gutenberg-Universit\"at Mainz, 55099 Mainz, Germany}
\affiliation{Helmholtz-Institut Mainz, Johannes Gutenberg-Universit\"at Mainz, 55099 Mainz, Germany}
\affiliation{GSI Helmholtzzentrum f\"ur Schwerionenforschung GmbH, Planckstra{\ss}e 1, 64291 Darmstadt, Germany}

\collaboration{for the Baryon Scattering (BaSc) Collaboration}

\date{\today}

\begin{abstract}

Understanding how the properties of the visible Universe depend on the parameters of the Standard Model of particle physics requires \emph{ab initio} theoretical computations. To that end, we investigate the quark-mass dependence of low-energy nucleon-nucleon scattering in the deuteron and dineutron channels. We perform lattice-QCD calculations at pion masses $M_\pi\approx 200$ and 280~MeV and compare their results to phenomenology at the physical point and to a previous calculation at $M_\pi\approx 710$~MeV. Scattering amplitudes, phase shifts, and threshold parameters are determined from finite-volume spectra using quantization conditions that explicitly incorporate long-distance effects from one-pion exchange. We analyze the pion-mass dependence of scattering lengths and effective ranges using chiral effective field theory with Weinberg and Kaplan-Savage-Wise power countings. From this, we estimate the pion-mass derivatives of the inverse scattering lengths and binding momentum, at the physical point. Both power countings indicate that the deuteron unbinds near $M_\pi\approx 220$~MeV.
\begin{tikzpicture}[remember picture,overlay,shift={(current page.north west)}]
  \node[anchor=north west,xshift=0cm,yshift=0cm]
  {\href{https://basc-collaboration.github.io/}{\includegraphics[width=1.8cm]{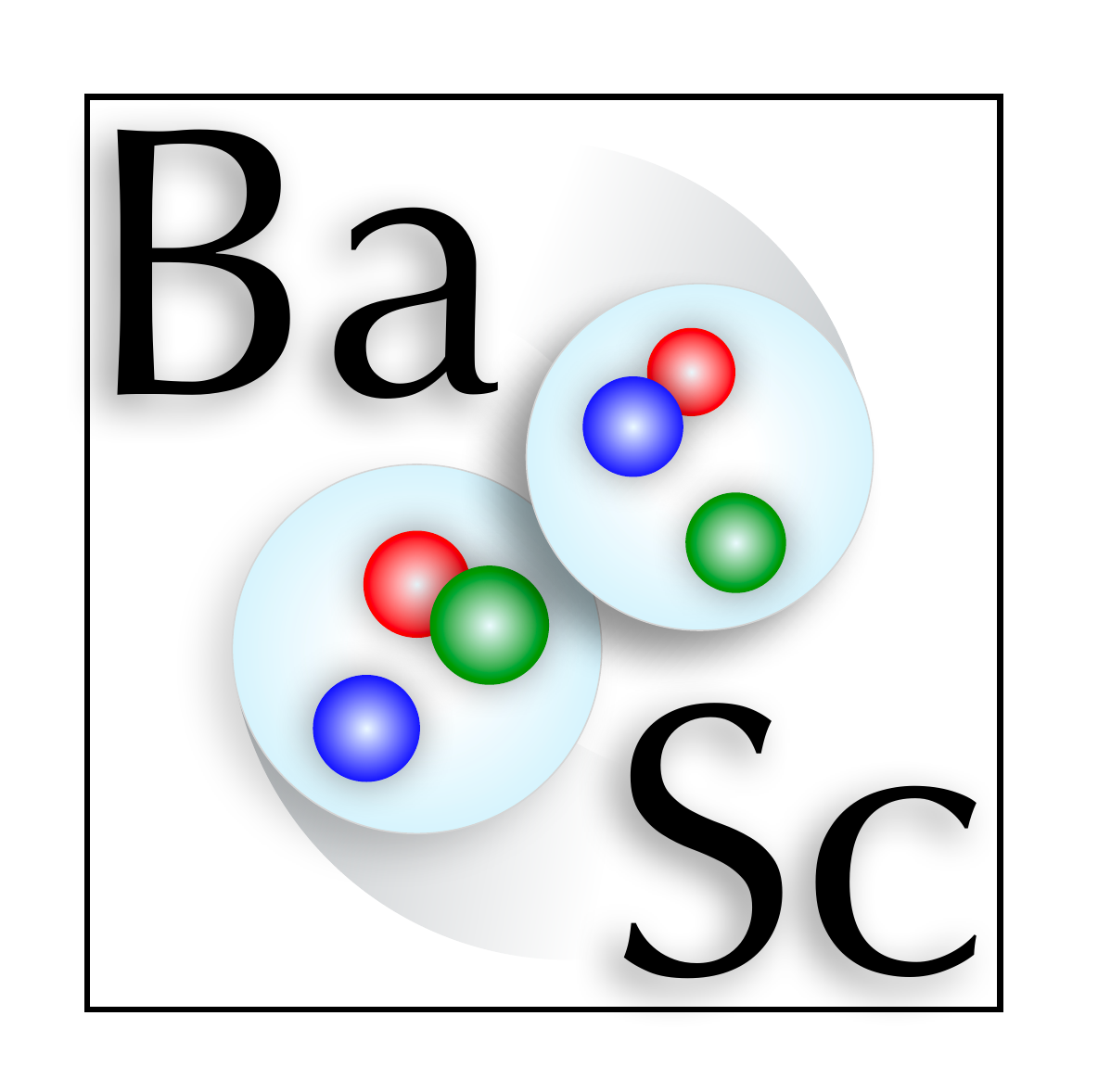}}};
no \end{tikzpicture}
\end{abstract}

\maketitle

\hd{Nuclear physics from first principles}
It is widely accepted that nuclear physics, comprising properties and interactions of nucleons and nuclei, emerges from the Standard Model and, in particular, from quantum chromodynamics (QCD), the theory of the strong interaction.
Nevertheless, the nonperturbative nature of low-energy QCD and the complexity of nuclear many-body systems limit the range of nuclear phenomena that can be studied from first-principles calculations~\cite{Beane:2010em,Detmold:2019ghl,Drischler:2019xuo,Davoudi:2020ngi,Cirigliano:2020yhp}. 

A deeper understanding of nuclear systems can have implications spanning a broad range of scales, from few-neutron interactions and electroweak processes in nuclei to the dense matter found in neutron stars~\cite{Beane:2010em,Detmold:2019ghl,Drischler:2019xuo,Davoudi:2020ngi,Fore:2023gwv,Cirigliano:2024ocg}. 
Even an improved description of the simplest nuclear system, that of two nucleons, can have repercussions for cosmology via Big Bang nucleosynthesis. Its onset was delayed by the deuterium bottleneck arising from the small deuteron binding energy; thus, the sensitivity of this binding to quark masses is a key input for constraining variations of Standard Model parameters between the epoch of nucleosynthesis and the present~\cite{Beane:2002vs,Epelbaum:2002gb,Bedaque:2010hr,Berengut:2013nh,Hall:2014dfa,Burns:2024ods, Meyer:2024auq}.

Lattice QCD has emerged as a powerful tool for building a first-principles understanding of nuclear physics~\cite{Detmold:2019ghl,Drischler:2019xuo,Cirigliano:2020yhp,Davoudi:2020ngi}. Several single-nucleon observables can now be determined with controlled uncertainties~\cite{FlavourLatticeAveragingGroupFLAG:2024oxs}, but two-nucleon observables remain challenging. Following more than two decades of methodological and computational advances~\cite{Beane:2010em,Drischler:2019xuo,Luscher:1986pf,Luscher:1990ck,HadronSpectrum:2009krc,Beane:2006mx,Ishii:2006ec,Ishii:2012ssm,Morningstar:2011ka}, recent calculations using large correlation matrices with many interpolating operators have provided increasingly precise constraints on two-nucleon scattering amplitudes at heavier-than-physical quark masses~\cite{Horz:2020zvv,Amarasinghe:2021lqa,Detmold:2026vjx,Detmold:2026shh,Detmold:2024iwz,Green:2022rjj,BaryonScattering:2025ziz,Zhang:2026cco,Green:2021qol}. Extending these calculations to lighter quark masses is the next critical step, enabling matching to effective field theories (EFTs), which could then be used for a more predictive description of few-nucleon systems.

Approaching physical quark masses presents several challenges. First, the signal-to-noise problem in multi-nucleon correlation functions becomes severe~\cite{Parisi:1983ae,Lepage:1989hd}. Second, as the pion mass decreases, long-distance interactions generated by one-pion exchange become increasingly important and must be taken into account for the analysis of $NN$ interactions. 
Advances in computational resources, algorithms, and correlation-function construction help mitigate the former challenge. The latter has motivated new finite-volume approaches that explicitly incorporate long-range interactions, including the plane-wave quantization condition and relativistic formulations that include effects from the left-hand cut~\cite{Meng:2021uhz,Raposo:2023oru,Raposo:2025dkb,Dawid:2024oey,Bubna:2024izx}.

In this Letter, we determine the quark-mass dependence of low-energy \(NN\) scattering amplitudes using lattice QCD calculations at pion masses \(M_\pi \approx 200\) and \(280\)~MeV, and compare them to phenomenological values at physical pion mass and to results from a previous lattice calculation at $M_\pi\approx 710$ MeV~\cite{BaryonScattering:2025ziz}. We study both the isospin-0 (deuteron) and isospin-1 (dineutron) channels, particularly the coupled \(^{3}S_{1}\)--\(^{3}D_{1}\) and uncoupled \(^{1}S_{0}\) partial waves, respectively, using the spectroscopic ${}^{2S+1}L_J$ notation. We compute the finite-volume energy spectra and analyze them using two of the new finite-volume formalisms.
From these spectra, we extract the phase shifts, threshold parameters, and near-threshold pole positions. We then analyze the pion-mass dependence of the results with a nuclear EFT, using two different power countings from Weinberg~\cite{Weinberg:1990rz,Weinberg:1991um} and from Kaplan-Savage-Wise (KSW)~\cite{Kaplan:1996xu,Kaplan:1998tg,Kaplan:1998we}. The results constrain for the first time the evolution of scattering parameters and pole positions as the quark masses are varied. This Letter is accompanied by a longer article that provides the full details of the lattice calculation and amplitude analysis~\cite{BaryonScattering:2026voj}. Here, we present the main results and focus on their physical interpretation and their description within effective field theory.

\hd{Lattice QCD setup}
The lattice calculations of this Letter use two $N_f=2+1$ ensembles from the Coordinated Lattice Simulations consortium with nonperturbatively $O(a)$-improved Wilson fermions~\cite{Bruno:2014jqa} and pion masses of approximately $200$ and $280$~MeV.  We also reanalyze the lattice results from \rcite{BaryonScattering:2025ziz} at a heavier pion mass. The parameters of these ensembles and relevant single-hadron masses are summarized in \Cref{tab:ensems}, and all ensembles satisfy $M_\pi L>4$.

Single- and two-hadron correlation functions are constructed using distillation~\cite{HadronSpectrum:2009krc}. 
Two-nucleon operators are built from momentum-projected single-nucleon operators and organized into sectors of fixed isospin, total momentum, and irreducible representation of the cubic symmetry little group. Within each sector, we construct large Hermitian correlation matrices and solve a generalized eigenvalue problem to form rotated correlators that approximately isolate the finite-volume energy levels.

Each rotated two-nucleon correlator is fitted simultaneously with the corresponding single-nucleon reference correlators using Bayesian multiexponential fits~\cite{BaryonScattering:2025ziz}. The two-nucleon energy is parametrized as
\begin{equation}
    E^{(NN)} = E^{N_1} + E^{N_2} + \delta E,
    \label{eq:ENN}
\end{equation}
where $E^{N_1}$ and $E^{N_2}$ are the lattice single-nucleon energies at the reference momenta, and $\delta E$ is the finite-volume energy shift, which contains information about the interactions. The simultaneous fits directly constrain $\delta E$, exploiting correlations between the single- and two-nucleon correlators, and including additional exponentials to account for residual excited-state contributions. We assess stability under variations of the fit range and number of exponentials, and use two independent spectrum-extraction strategies to estimate the associated systematic uncertainty. The resulting energy shifts are used as inputs to constrain the scattering amplitudes through the finite-volume formalisms.

\begin{figure*}
     \centering
     \includegraphics[width=\textwidth]{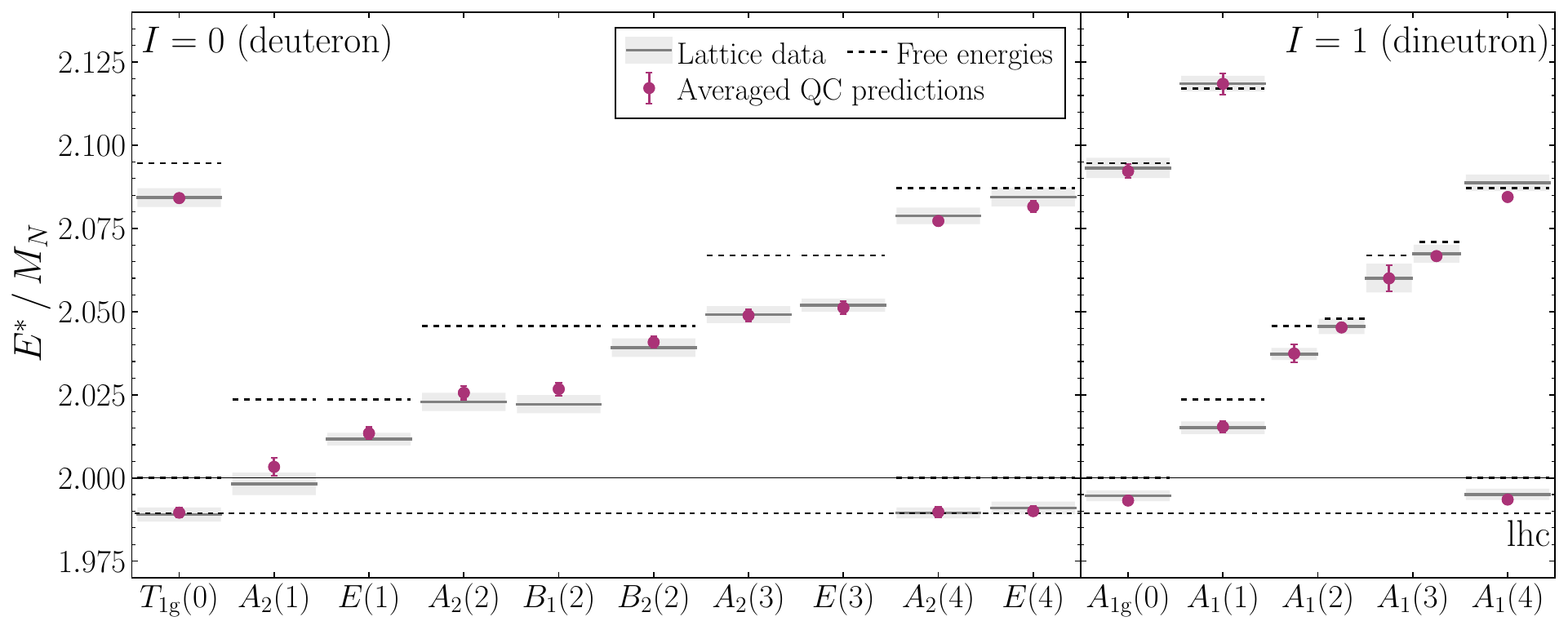}
     \caption{\label{fig:J250energyfit} 
     Summary of the spectrum determination and the fits to the quantization conditions on the ensemble J250 with $M_\pi \approx 200$ MeV for both isospin channels and irreps(frames). The position of the left-hand cut is indicated as a horizontal dashed line.
     }
\end{figure*}

\begin{table}[t]
%  \centering
\caption{\label{tab:ensems}
Lattice ensembles used in this work. Listed are the lattice geometry, pion, kaon, and nucleon masses in MeV, and lattice spacing in fm. All ensembles employ periodic boundary conditions.
}
\begin{ruledtabular}
  \begin{tabular}{c c c c c c}
%    \toprule
    Name
    & $(L/a)^3 \times (T/a)$
    & $M_\pi$
    & $M_K$
    & $M_N$
    & $a$ \\
    \midrule
    C103 & $48^3 \times 96$ & 709(8) & 709(8) & 1603(19) & 0.0865(10) \\
    D251 & $64^3 \times 128$ & 283(3) & 462(5) & 1039(12) & 0.0642(7)  \\
    J250 & $64^3 \times 192$ & 199(2) & 501(6) & 969(11)  & 0.0642(7)  \\
%    \bottomrule
  \end{tabular}
\end{ruledtabular}
\end{table}

\hd{Two-nucleon scattering amplitudes from lattice QCD}
Following L\"uscher's approach~\cite{Luscher:1986pf,Luscher:1990ck,Luscher:1990ux,Rummukainen:1995vs,Kim:2005gf,Briceno:2013lba,Briceno:2014oea,Briceno:2013bda}, we use the finite-volume energy spectrum to constrain scattering amplitudes. 
However, while the use of the two-particle quantization condition (QC) for two-nucleon systems is well established at heavy quark masses~\cite{BaryonScattering:2025ziz}, at lighter masses the formalism must be modified to account for the effects of one-pion exchange. Specifically, the partial-wave amplitudes have a left-hand cut beginning at the branch point $p^{2}=-M_\pi^2/4$ and extending to lower $p^{2}$, where $p$ is the center-of-mass momentum. As the pion mass decreases, this branch point approaches threshold and can overlap the subthreshold region containing finite-volume states~\cite{Green:2021qol}. On the cut, the standard quantization condition is no longer applicable, while, just above it, energies can receive enhanced exponentially suppressed finite-volume corrections. The presence of the left-hand cut also limits the range over which higher partial waves are suppressed.

To overcome these complications, we employ two approaches that incorporate one-pion exchange explicitly (see Refs.~\cite{Bubna:2024izx,Bubna:2025gsd,Dawid:2024oey,Rodas:2026zmm} for other approaches). The plane-wave quantization condition (QC)~\cite{Meng:2021uhz} formulates the scattering problem in a finite-volume momentum basis, with an interaction potential containing one-pion exchange and short-range contact terms. 
The corresponding QC follows from the finite-volume Lippmann-Schwinger equation. By contrast, the relativistic left-hand-cut formalism~\cite{Raposo:2023oru,Raposo:2025dkb} isolates the one-pion exchange contribution and constrains a short-range K-matrix-like quantity through a modified Lüscher-like QC. In both approaches, the finite-volume momentum sums are regulated with a cutoff. These two formalisms not only make it possible to incorporate one-pion exchange effects, but also allow one to simultaneously analyze different isospin channels with a common pion-nucleon coupling, and to capture all partial waves generated by the long-range interactions. The equivalence of the two formalisms used in this work has been demonstrated in the nonrelativistic regime for the similar $DD^*$ system~\cite{Dawid:2025wsn}.

After the pion-nucleon coupling and either the short-range part of the interaction potential or the modified K matrix have been constrained, they can be used to reconstruct the physical amplitudes by solving the corresponding infinite-volume integral equations.

We analyze the spectra using both formalisms for the two ensembles with lighter pion mass. For the heavier ensemble from~\cite{BaryonScattering:2025ziz}, only the plane-wave QC is used for the reanalysis, which has some advantages compared to the traditional Lüscher QC even at this heavy pion mass.  
We perform a range of fits to assess systematic uncertainties associated with the spectrum determination, the parametrization of the short-range K matrix or potential, and the regulator cutoff. All these variations, together with the comparison between the two formalisms, enter our assessment of systematic uncertainties, and are discussed in detail in the companion paper~\cite{BaryonScattering:2026voj}. 

The comparison between the lattice-determined energies and the finite-volume predictions is shown in \Cref{fig:J250energyfit} for J250, the ensemble with the lightest pion mass. The lattice results are averaged with equal weights over the two independent spectrum determinations, including systematic uncertainty. The predictions are similarly averaged over all fits, assigning equal weights to each combination of formalism and spectrum determination and using the Akaike Information Criterion  \cite{Akaike1,Jay:2020jkz,Frison:2023lwb,Neil:2022joj,Neil:2023pgt} weights within each formalism/spectrum combination. Similar averaging is also used for other scattering quantities presented later. Broad agreement between lattice and fit results is observed, and a substantial fraction of the fits yields a $\chi^2$ per degree of freedom (dof) close to or below unity.

\begin{figure}
    \centering
    \includegraphics[width=\columnwidth]{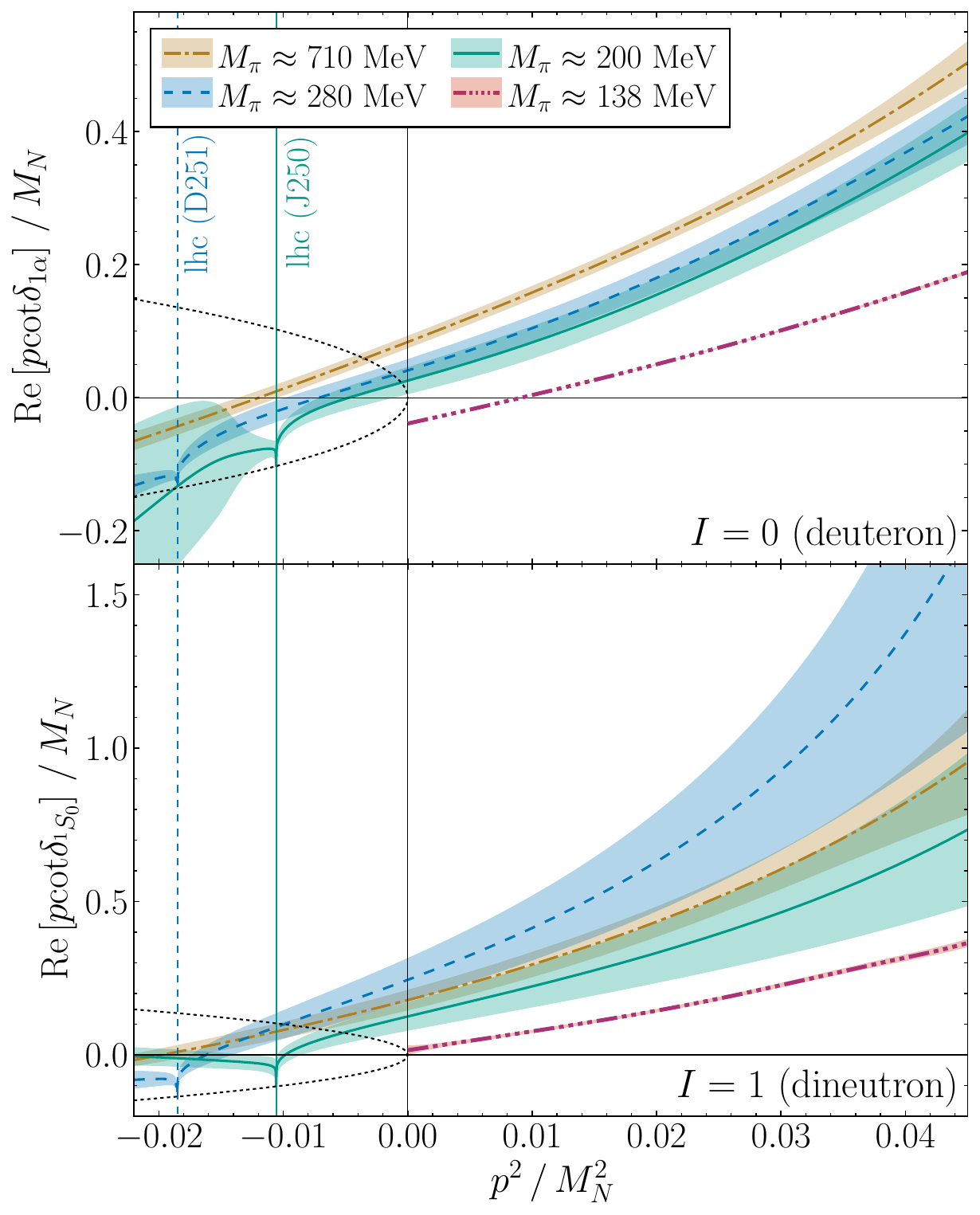}
    \caption{\label{fig:kcotI0I1withphysical} Phase shift, shown as the real part of $p\cot\delta$, versus scattering momentum, in nucleon-mass units, for the lowest partial wave in the deuteron ($\alpha$ wave for $J^P=1^+$, top) and dineutron (${}^1S_0$ wave, bottom) channels. We show all ensembles considered in this work, together with phenomenological curves at the physical pion mass. The mean value and error of the latter are assigned from the spread of various analyses of nuclear $NN$ phase shifts~\cite{nnonline,Stoks:1993tb,Stoks:1994wp,Rijken:1995pu,Rijken:1995pw}, with an additional systematic error assigned to $I=1$ related to the isospin-breaking differences between proton-proton and neutron-proton data~\cite{Henley:1966zza,Miller:2006tv}. Dotted black lines represent the bound- (bottom branch) and virtual-state (top branch) pole conditions, given in \cref{eq:boundstate}.}
\end{figure} 

\hd{Quark-mass dependence of the phase shift}
Phase shifts are determined on all three ensembles after solving the integral equations associated with each QC, allowing for a direct comparison of their quark-mass dependence. Such a comparison for the energy-dependent phase shifts is shown in \Cref{fig:kcotI0I1withphysical} for the lowest partial waves present in the two isospin channels studied here, together with the phenomenological phase shifts from experiment~\cite{nnonline,Stoks:1993tb,Stoks:1994wp,Rijken:1995pu,Rijken:1995pw}. For the dineutron channel, we consider the ${}^1S_0$ wave, while for the deuteron channel we present the $\alpha$ eigenphase in the Blatt-Biedenharn (BB) decomposition~\cite{Blatt:1952zza} of the coupled $J^P=1^+$ sector. 
In the case of the deuteron channel, expressed in $M_N$ units, there is a clear trend of $p\cot\delta$ to monotonically approach the experimental result, while the dineutron channel only displays monotonic behavior for the lighter-pion-mass results.

Of particular interest is the behavior of the cotangent of the phase shift close to threshold, and in particular the value of the threshold parameters, defined from the effective range expansion (ERE),
\begin{equation}
    p\cot\delta
    = -\frac{1}{a_0} + \frac{1}{2}r_0p^2 + P_0p^4
    + \mathcal{O}(p^6)\,,
\end{equation}
where $\delta$ denotes the ${}^1S_0$ phase shift ($I=1$) or the $\alpha$ eigenphase ($I=0$), while $a_0$ is the scattering length, $r_0$ is the effective range and $P_0$ is the shape parameter. The averaged results for all these parameters determined from our lattice analysis are presented in the End Matter in \cref{tab:finalScatteringParameters}. Phenomenological values at the physical pion mass are also summarized, where the assigned errors include systematics from isospin-breaking effects---see \cref{app:physicalpoint} for a detailed description of how these are assessed. 

The $NN$ scattering amplitude can contain poles at values of the momentum $p$ that satisfy
\begin{equation}\label{eq:boundstate}
    p\cot\delta-ip=0\,.
\end{equation}
For bound or virtual states, the pole momentum can be written as $p=i\kappa$, where $\kappa>0$ corresponds to a bound state and $\kappa<0$ to a virtual state. The signed binding momentum $\kappa$ is reported in the last column of \cref{tab:finalScatteringParameters}. The corresponding pole energy relative to threshold is $B=2M_N-2\sqrt{M_N^2-\kappa^2}$.

The negative inverse scattering length $-1/a_0$, which corresponds to the intercept of $p\cot\delta$, can be associated to the presence of a pole close to threshold for small values. Generally, a negative intercept is associated with a bound state, whereas a positive intercept is associated with a virtual state. The central values of all our determinations yield positive intercepts. However, in the $I=0$ channel at $M_\pi\approx200$~MeV, the intercept is compatible with zero within $1.3\sigma$, indicating proximity to the unitary limit and the virtual-to-bound-state transition.

\hd{EFT analysis}
To capture the pion-mass dependence of the nucleon-nucleon scattering parameters, we match our results to EFTs. The EFT description of $NN$ scattering has been extensively studied~\cite{Epelbaum:2008ga, Hammer:2019poc}. A central ingredient is the power counting, which organizes contributions according to their expected size in the low-energy expansion. Different schemes have been proposed to explain the unnaturally large values of the scattering lengths. Here, we consider the Weinberg power counting~\cite{Weinberg:1990rz,Weinberg:1991um} and the KSW power counting~\cite{Kaplan:1996xu, Kaplan:1998tg, Kaplan:1998we}, and focus only on the light-quark-mass dependence; the effect of the slight difference in the strange-quark mass is assumed to be negligible.

In the Weinberg scheme, the perturbative  expansion is applied to the interaction potential, which is truncated at a given order and then resummed to all orders through the Lippmann-Schwinger equation to obtain the scattering amplitude. At leading order (LO), both one-pion exchange and zero-derivative contact terms are therefore treated nonperturbatively. Higher-order contributions, including the two-derivative contact terms introduced at next-to-leading order (NLO), are likewise included in the iterated potential.
On the other hand, the KSW scheme resums only the leading contact interaction to account for the large scattering lengths, while NLO contributions from pion exchange and subleading contact interactions are treated perturbatively. The expansion in KSW counting is therefore organized directly at the amplitude level.

\begin{figure}
    \centering
    \includegraphics[width=0.48\textwidth]{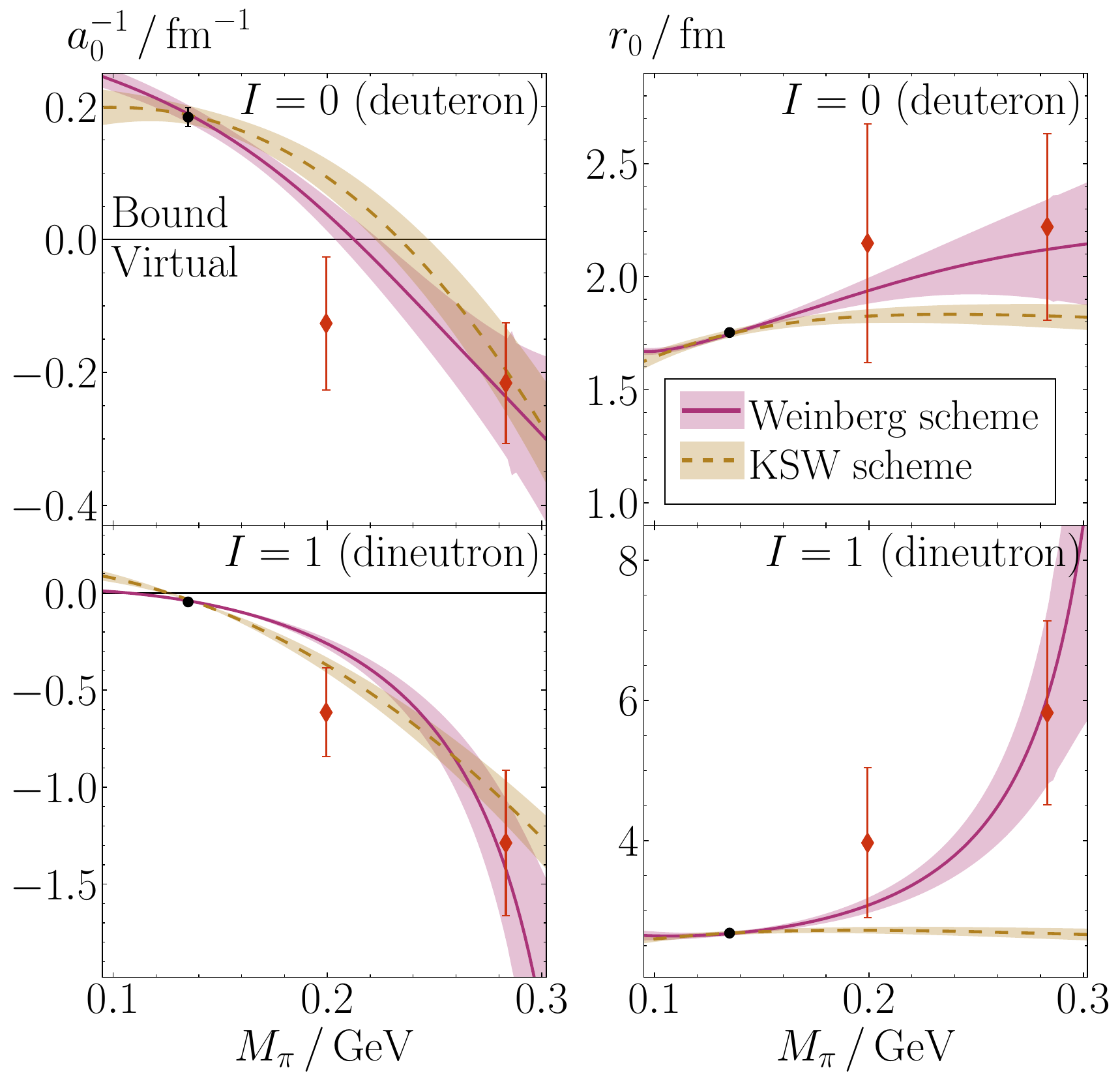}
    \caption{\label{fig:pionmassdependence} Inverse scattering length (left) and effective range (right) versus pion mass, for the deuteron (top) and dineutron (bottom) channels. Phenomenological values are represented as black dots. EFT predictions are presented as magenta and yellow bands, with each corresponding to a different power counting.
     }
\end{figure}

We perform an EFT analysis of the scattering lengths and effective ranges using both  Weinberg and KSW power countings through NLO. Phenomenological values for physical-point scattering parameters are taken as input, and we use our lattice results from the two lightest ensembles to constrain the pion-mass dependence. The resulting EFT descriptions predict how the scattering parameters evolve away from the physical point. 
Further details of the procedure and the explicit expressions are provided in the End Matter in \cref{app:EFT}, and  a description of the fit methodology is discussed in \cref{app:EFTmethodology}, which also summarizes the fit results.

EFT predictions for both $a_0^{-1}$ and $r_0$ are presented in \Cref{fig:pionmassdependence}, where the dependence of the nucleon mass on the pion mass is determined from fitting to the lattice results; see \cref{app:EFTmethodology}. This figure also includes the physical-point data and the lattice results that are used for the fits.  The two power countings yield qualitatively consistent results, although in some regions there is more than $1\sigma$ tension. The strongest disagreement is found in the $I=1$ channel, where the Weinberg scheme fails to describe the scattering length at $M_\pi\approx 200$ MeV, while the KSW analysis cannot reproduce the large values of the lattice results for $r_0$. The observed tensions are an indication of EFT limitations at the given order.

The EFT fits constrain the low-energy couplings of the effective theory, which are generally scheme dependent and are not themselves observables.  We therefore report quantities derived from the physical amplitudes: the derivatives of the inverse scattering length and of the binding momentum, with respect to the pion mass, evaluated at the physical point, and the unitary pion mass $M_\pi^\star$, defined by $a_0^{-1}(M_\pi^\star)=0$.  The results are summarized in \Cref{tab:EFTobservables}. Both schemes place the deuteron virtual-to-bound-state transition near $M_\pi^\star\approx220$~MeV, slightly above the lightest simulated pion mass. The tension with the lattice central values may reflect statistical fluctuations, unaccounted-for systematic uncertainties such as discretization effects, or higher-order EFT contributions. For the dineutron, both central values place the transition below the physical pion mass, although as with other $I=1$ observables, there is significant tension between the two schemes.

\begin{table}[t]
\caption{\label{tab:EFTobservables}
Quantities derived from the EFT fits using KSW and Weinberg power counting. Derivatives are evaluated at the physical pion mass. The binding-momentum derivative is dimensionless, with $\kappa$ and $M_\pi$ expressed in the same units.}
\begin{ruledtabular}
\begin{tabular}{c c c c c}
     Isospin & Scheme & $\text{d} a_0^{-1}/\text{d} M_\pi$  &  $\text{d} \kappa/\text{d} M_\pi$ & $M_\pi^*\,/\,\mathrm{MeV}$ \\
    \midrule
     \multirow{2}{*}{0}  & Weinberg  & $-0.35(09)$ & $-0.56(10)$  & $213(13)$ \\
     %$-0.349(88)$ \  & $-0.56(10)$  & $213(13)$ \\
 & KSW & $-0.16(11)$ &$-0.24(19)$  & $228(13)$ \\\midrule
  \multirow{2}{*}{1}  & Weinberg  & $-0.39(06)$  & $-0.35(05)$  & $108(06)$ \\
  %$-0.394(58)$  & $-0.353(51)$  & $107.9(6.1)$ \\
 & KSW &$-0.85(15)$ &$-0.77(14)$ & $127(02)$\\    
%& KSW &$-0.85(15)$ &$-0.77(14)$ & $126.8(2.4)$\\    
%\bottomrule
\end{tabular}
\end{ruledtabular}    
\end{table}

\hd{Conclusion and outlook}
This Letter presents the first determination of $NN$ scattering amplitudes at pion masses of $M_\pi\approx200$--$280$~MeV, i.e.\ in a pion-mass range where contact with effective theories is possible. A key novelty of this analysis is the usage of finite-volume formalisms that account for the finite-volume effects of the left-hand cut induced by one-pion exchange, which in addition allows us to analyze both isospin channels simultaneously with a common pion-nucleon coupling. 

In both isospin channels, the phase shift and threshold parameters qualitatively evolve toward the known phenomenological values as the pion mass decreases. More quantitative analyses of the scattering parameters using KSW and Weinberg power counting provide a description of this evolution and place the deuteron virtual-to-bound-state transition near $M_\pi\approx 220$~MeV.

Discretization effects remain the principal outstanding systematic uncertainty and require calculations at additional lattice spacings. Additional pion masses are also needed to resolve the mass dependence and further constrain the EFT extrapolations. This work establishes a path forward for combined lattice-QCD and EFT studies of few-nucleon systems with controlled systematics.

{\bf Acknowledgements}:
We acknowledge helpful discussions with Vadim Baru, Ra\'ul Brice\~no, Sebastian Dawid, Evgeny Epelbaum, Andrew Jackura, Gerald A. Miller, and Nelson Pitanga Lachini. JBB thanks the University of Bern and the Albert Einstein Center for Fundamental Physics for their hospitality during his visit.

The work of FRL and MS was supported in part by the Swiss
National Science Foundation (SNSF) through grant No. 200021-236432, and the Platform for
Advanced Scientific Computing (PASC) project ``ALPENGLUE''. 
The work of ABR, JB, DL is supported by the European Research Council (ERC) consolidator grant StrangeScatt-101088506.
The work of AN is supported by the NSF Faculty Early Career Development Program (CAREER) under award PHY-2047185.
The work of TRR was supported by the NSF through cooperative agreement 2020275. CM acknowledges support from the NSF under awards PHY-2514831, OAC-2311430, and OAC-2608348.
This work was supported in part by the US Department of Energy, Office of Science under grant contract numbers DE-AC02-05CH11231 (ML, TRR, AWL), through the
“Nuclear Theory for New Physics” Topical Collaboration award No.\ DE-SC0023663 (ML, TRR, AWL).
The work of HW was supported by Deutsche Forschungsgemeinschaft
(German Research Foundation, DFG) through the Cluster of Excellence “Precision Physics, Fundamental Interactions and Structure of Matter” (PRISMA++ EXC 2118/2), funded within the German Excellence
Strategy (Project No.\ 390831469). DM is funded by the Heisenberg Programme of the Deutsche Forschungsgemeinschaft (DFG, German Research Foundation) project number 454605793.

The authors gratefully acknowledge the computing time provided by HLRS on ``Hawk'' through the project ``BarScattChi'', which was used to generate the J250 ensemble with openQCD~\cite{openQCD} that was seeded through resources of the National Energy Research Scientific Computing Center (NERSC), a Department of Energy User Facility using NERSC award ERCAP0024058. This work presents the first use of this ensemble.

The perambulators were computed using either
QUDA-LapH~\cite{Clark:2025cuz}, which is built on
QUDA~\cite{Clark:2009wm, Babich:2011np, Clark:2016rdz, Clark:2017ekr,
  Clark:2025cuz}, or a combination of QDP++~\cite{Edwards:2004sx} and
QUDA. The mode triplets necessary for contractions involving baryons
were computed using QDP++ and PRIMME~\cite{PRIMME}. The contractions
were performed using NumPy~\cite{numpy} and CuPy~\cite{cupy}.

Part of the analysis was performed on UBELIX (\url{https://www.id.unibe.ch/hpc}), the HPC cluster at the University of Bern, and on grants from the Swiss National Supercomputing Centre (CSCS) under project IDs \texttt{c36} and \texttt{lp151} on Alps, as well as on the HPC cluster Elysium
of the Ruhr-Universit\"at Bochum, subsidized by the DFG(INST213/1055-1). Some of the spectrum fits were performed in the PAX cluster at DESY Zeuthen. The perambulators on D251 used a grant from the Swiss National Supercomputing Centre (CSCS) under project ID \texttt{lp53} on Alps. The perambulators on ensemble J250 and the contractions for ensembles
J250 and D251 were computed using resources provided by the Gauss
Centre for Supercomputing e.V.\ (\url{www.gauss-centre.eu}) on
JUWELS~\cite{juwels} at Jülich Supercomputing Centre.
We are grateful to our colleagues within the CLS initiative for sharing the D251 ensemble.

\clearpage
\appendix
\crefalias{section}{appendix}
\widetext

\begin{center}
    \large\textbf{End Matter}
\end{center}

\section{Expressions for the EFT analysis}
\label{app:EFT}
 
\nopagebreak
 
In the KSW scheme, the perturbative expansion is applied to the amplitude, $\mathcal A(p) = \mathcal A_{-1}(p) + \mathcal A_0(p) + \cdots$, 
with expressions for the LO ($\mathcal A_{-1}$),
and NLO ($\mathcal A_{0}$), amplitudes provided in Refs.~\cite{Kaplan:1998tg, Kaplan:1998we}.
The coefficients of the ERE are determined by solving for the K matrix and expanding the inverse amplitude in powers of $\mathcal A_0/\mathcal A_{-1}$.
The expressions for $1/a_0$ and $r_0$ were originally presented in the power-divergence subtraction (PDS) scheme with the choice $\mu = M_\pi$~\cite{Kaplan:1998tg, Kaplan:1998we}.
Here, we give the general expressions, 
which have no isospin dependence aside from the LEC values:
\begin{align}
\frac{1}{a_0} & = 
\frac{4 \pi}{M_N C_0} \left[ 
    1 + \frac{M_N C_0}{4 \pi} \mu_{\rm PDS} 
    - \frac{M_\pi^2 D_2}{C_0} 
    + \frac{g_{\pi NN}^2}{8 \pi}\left[ \frac{\mu_{\rm PDS} - M_\pi}{M_N} 
    + \frac{C_0}{8 \pi} \left( 
        (M_\pi - \mu_{\rm PDS})^2 + \frac{M_\pi^2}{2} \log \frac{\mu^2}{M_\pi^2} 
    \right) \right]
    \right]\, ,
\\
r_0 & = \frac{g_{\pi NN}^2}{8 \pi M_N} \left[ 
    1 
    - \frac{32 \pi}{3 M_\pi M_N C_0} 
    + \frac{32 \pi^2}{M_\pi^2 M_N^2 C_0^2} 
    + \mu_{\rm PDS} \left( \frac{16 \pi}{M_\pi^2 M_N C_0} - \frac{8}{3 M_\pi} \right) 
    + \frac{2 \mu_{\rm PDS}^2}{M_\pi^2} \right] 
    + \frac{8 \pi C_2}{M_N C_0^{2}} \, , 
\end{align}
In the above expressions, one sets $\mu_{\rm PDS}=\mu$ for the PDS scheme and $\mu_{\rm PDS}=0$ for the $\overline{\rm MS}$ scheme. 
In PDS, $C_0$, $C_2$ and $D_2$ depend on $\mu$. 
In $\overline{\rm MS}$, only $D_2$ depends on $\mu$, and it cancels the logarithmic dependence on $\mu$ in $1/a_0$, while $r_0$ is independent of $\mu$.
\bigskip

We also use the Weinberg scheme~\cite{Weinberg:1990rz,Weinberg:1991um}.
%The second EFT scheme we follow is that of . 
In this case, the full two-particle scattering amplitude is determined from the partial-wave-projected Lippmann-Schwinger equation,
given in terms of the partial-wave projected regularized EFT potential,  $V$, and for which we use a semi-relativistic kernel based on the Blankenbecler-Sugar equation~\cite{Blankenbecler:1965gx, Meng:2023bmz}. In this work, we use a potential containing NLO contact terms and a LO one-pion-exchange part, $V=V_\text{cont}+V_\text{OPE}$, analogous to the one used in the lattice analysis performed with the plane-wave QC~\cite{BaryonScattering:2026voj}. Near threshold, effects from the two-pion exchange can be absorbed into the contact interactions perturbatively~\cite{Berengut:2013nh}. The contact interactions read:
\begin{align}
    V^{00}_\text{cont}(p',p) & = \left[\tilde{C}_{{}^1S_0}+\tilde{D}_{{}^1S_0}\delta+{C}_{{}^1S_0}(p'^2 + p^2)\right]\times H_\text{cut}(p',p),\\[5pt]
    V^{11}_\text{cont}(p',p) & = \begin{pmatrix} 
       \tilde{C}_{{}^3S_1}+\tilde{D}_{{}^3S_1}\delta+{C}_{{}^3S_1}(p'^2 + p^2) & {C}_{{}^3S_1-{}^3D_1}p^2 \\
       {C}_{{}^3S_1-{}^3D_1}p'^2 & 0
    \end{pmatrix}\times H_\text{cut}(p',p),
\end{align}
where matrix rows and columns correspond to different orbital angular momentum, and the potential superscripts indicate the total and spin angular momentum, in this same order. Also, we define $\delta=(M_\pi^2-M_{\pi,\text{phys}}^2) / M_{\pi,\text{phys}}^2$ and $H_\text{cut}(p',p)=\exp[-(p'^2+p^2)/\Lambda^2]$ the cutoff function. For the regulated OPE potential we use~\cite{Reinert:2017usi}: 
\begin{equation}
    V_\text{OPE}(\boldsymbol{p}',\boldsymbol{p})=-c_I (g_{\pi NN}^2+\delta\tilde{D}_g)\biggl\{\frac{[(\boldsymbol{p}'-\boldsymbol{p})\cdot\boldsymbol{\sigma}_1][(\boldsymbol{p}'-\boldsymbol{p})\cdot\boldsymbol{\sigma}_2]}{(\boldsymbol{p}'-\boldsymbol{p})^2+M_\pi^2}+C(M_\pi,\Lambda)\boldsymbol{\sigma}_1\cdot\boldsymbol{\sigma}_2\biggr\}\times\text{exp}\left[-\frac{(\boldsymbol{p}'-\boldsymbol{p})^2+M_\pi^2}{\Lambda^2}\right]
\end{equation}
with $c_I=1$ (dineutron) or $-3$ (deuteron), $\boldsymbol{\sigma}_{1,2}$ a vector of Pauli matrices acting on the nucleon spins and $C(M_\pi,\Lambda)$ given in Eq.~(50) of \rcite{BaryonScattering:2026voj}. We keep the cutoff identical for the contact and OPE terms.  
We extract the threshold observables from the determined phase shifts: ${}^1S_0$ for $I=1$, and the $\alpha$-wave in the BB decomposition for $I=0$.

\section{\label{app:physicalpoint} Isospin-symmetric physical-point inputs}

Our analysis requires estimates of the physical-point isospin-symmetric quantities. This is important for low-energy $NN$ scattering: the fine tuning responsible for the large scattering lengths enhances their sensitivity to isospin-breaking effects~\cite{Henley:1966zza,Miller:2006tv}.  We take a conservative approach and assign large uncertainties for the scattering lengths. For the dineutron channel, we define $a_{{}^1S_0}^{\rm iso} = (a_{nn}+2a_{np})/3$ using the results from \rcite{Miller:2006tv}, and assign an error equal to the difference between this average and the $np$ channel; $pp$ results are not included into the average due to the larger electromagnetic effects. In the deuteron channel, we use the central value from the more recent \rcite{NavarroPerez:2014ovp}, and assign the same relative error as in the dineutron channel. For the effective ranges and shape parameters, which are not finely tuned, we take the results from the systematics row in table III of \rcite{NavarroPerez:2014ovp} for the dineutron channel, and from averaging the last nine rows of Table II in the same reference for the deuteron, corresponding to different descriptions of $np$ experimental data in the BB decomposition. For the latter, we assign an error equal to the standard deviation. The final values are presented in \cref{tab:finalScatteringParameters}, together with the model-averaged results obtained from lattice simulations; see the companion paper~\cite{BaryonScattering:2026voj} for details. Finally, we also take conservative estimates for the physical pion and nucleon masses, with $M_\pi^{\rm iso}\equiv 138.1(2.0)~\mathrm{MeV}$ and $M_N^{\rm iso} \equiv 938.9(0.7)~\mathrm{MeV}$.

{\squeezetable
\begin{table*}%[h]
\caption{\label{tab:finalScatteringParameters} 
Summary table with the scattering parameters determined from the lattice QCD calculations. The last row in each isospin block indicates the physical-point values from~\rrcite{Miller:2006tv,NavarroPerez:2014ovp}; see text for details. The values of $\kappa$ are determined from solving \cref{eq:boundstate} with a quartic expansion for $p\cot\delta$. 
}

\begin{ruledtabular}

\begin{tabular}{c c c c c c c c}
     Isospin & Ensemble & $M_\pi$ / MeV&  $a_0^{-1}$ / fm${}^{-1}$ & $a_0$ / fm &  $r_0$ / fm &  $P_0$ / fm${}^3$ & $\kappa$ / MeV \\
    \midrule
     \multirow{4}{*}{0} 
& C103 & 710  & $-0.685(78)$  & $-1.46(17)$  & $1.76(12)$  & $0.035(30)$  & $-95.1(8.3)$ \\
  & D251 & 280  & $-0.216(91)$  & $-4.9(3.0)$  & $2.22(41)$  & $0.26(39)$  & $-35(13)$ \\
  & J250 & 200  & $-0.13(10)$  & $-9(11)$  & $2.15(53)$  & $0.13(45)$  & $-22(16)$ \\
& \rcite{NavarroPerez:2014ovp} & 138.1(2.0) & 0.185(14) & 5.42(42) & 1.754(17) & 0.043(9) & 45.8(4.6)  \\
\midrule \multirow{4}{*}{1} 
& C103 & 710  & $-1.46(27)$  & $-0.69(13)$  & $2.59(28)$  & $0.169(45)$  & $-150(17)$ \\
  & D251 & 280  & $-1.29(38)$  & $-0.79(23)$  & $5.8(1.3)$  & $0.83(45)$  & $-103(13)$ \\
  & J250 & 200  & $-0.61(23)$  & $-1.65(61)$  & $4.0(1.1)$  & $-0.36(57)$  & $-70(13)$ \\
& \rrcite{Miller:2006tv,NavarroPerez:2014ovp} & 138.1(2.0) & $-0.0452(35)$ & -22.1(1.7) & 2.68(3) & $-0.48(2)$ & $-8.44(62)$  \\    
\end{tabular}

\end{ruledtabular}
\end{table*} 
}

\section{\label{app:EFTmethodology} Fit methodology and results}

We perform fits to $a_0^{-1}$ and $r_0$ for both isospins and three pion masses, namely the physical point from phenomenology and $M_\pi\approx 200$ and $280$ MeV from our lattice calculations. 
We account for correlations among the ERE parameters but neglect their correlations with the nucleon mass. We also neglect strange-quark effects throughout.

For the Weinberg power counting, we fit 9 parameters. For $g_{\pi NN}^2$, we use the physical value from \rcite{Reinert:2020mcu}, $(g_{\pi NN,\text{phys}})^2=175.7(1.9)$ as input, where the error includes systematic effects from the difference between the different isospin channels. This input is implemented via a prior, while other parameters are unpriored.  
We also include in the fit the $S$--$D$ mixing angle in the BB decomposition, $p^{-2} \varepsilon_1^\text{phys}|_{p=0}=7.63(10)\,\text{GeV}^{-2}$~\cite{NavarroPerez:2014ovp}, which we need to constrain the $C_{{}^3 S_1-{}^3D_1}$ contact term. Fits are repeated for several values of the cutoff parameter $\Lambda$. Results are outlined in \cref{table:fitresultsWeinberg}. Predictions for physical quantities presented in \cref{tab:EFTobservables} and \cref{fig:pionmassdependence} are obtained after model averaging with weights based on the AIC~\cite{Akaike1,Jay:2020jkz,Frison:2023lwb,Neil:2022joj,Neil:2023pgt}.

Fits with the KSW formula are performed simultaneously with fits to the pion-mass dependence of $M_N$~\cite{Bernard:2003rp},
\begin{equation}\label{eq:MN_NLO}
M_N = M_0 -4 c_1 M_\pi^2 
    -\frac{3\pi g_{\pi NN}^2 M_\pi^3}{32 \pi^2 M_0^2}\, ,
\end{equation}
with a total of 9 fit parameters, two of which appear only in the expression for the mass and $g_{\pi NN}$ is shared.  We list results for the $\overline{\rm MS}$ scheme at $\mu=300$~MeV for scattering parameters, as we find that the PDS scheme does not provide a reasonable description of the lattice data. 
We also observe that $\mathcal{A}_0/\mathcal{A}_{-1}$ is not small over the range of $135\lesssim M_\pi\lesssim280$~MeV, indicating further challenges with the EFT expansion.
The prior and posterior values of the fit parameters are listed in \cref{table:KSWfits}. The EFT prediction for the nucleon mass from this fit is used together with the fit results of both Weinberg and KSW schemes to produce the curves in \cref{fig:pionmassdependence} and determine the results in \cref{tab:EFTobservables}. 

{ \squeezetable
\begin{table}[h]
\caption{\label{table:fitresultsWeinberg}
Results for the EFT parameters in the Weinberg scheme. For these fits, $g_{\pi NN}$ is priored by its physical value $(g_{\pi NN}^{\text{phys}})^2=175.7(1.9)$, and all parameters presented other than $g_{\pi NN}^2$ and $\tilde{D}_g$ are divided by $10^3$ for readability. 
}

\begin{ruledtabular}    
    \begin{tabular}{c c c c c c c c c c c}

     $\Lambda\,/\,\mathrm{GeV}$ & $\chi_\text{aug}^2/\mathrm{dof}$ & $g_{\pi NN}^2$ & $\tilde{D}_g$ & $\tilde{C}_{{}^3S_1}$ & $\tilde{D}_{{}^3S_1}$  & $C_{{}^3S_1}\,/\, \text{GeV}^{-2}$ & $C_{{}^3 S_1-{}^3D_1}\,/\,\text{GeV}^{-2}$  &  $\tilde{C}_{{}^1S_0}$ &  $\tilde{D}_{{}^1S_0} $ &  $C_{{}^1S_0} \,/\, \text{GeV}^{-2}$\\
\midrule
    0.30 & 6.8/5  & $175.7(1.9)$  & $2(28)$  & $-4.87(14)$  & $0.68(17)$  & $-16.4(3.7)$  & $28(11)$  & $-4.048(86)$  & $0.88(10)$  & $-1.7(1.8)$ \\
0.35 & 5.9/5  & $175.7(1.9)$  & $-37(22)$  & $-4.27(10)$  & $0.36(14)$  & $-0.3(3.8)$  & $18(10)$  & $-3.915(91)$  & $0.747(97)$  & $8.7(1.7)$ \\
0.40 & 5.3/5  & $175.7(1.9)$  & $-49(20)$  & $-3.71(36)$  & $0.12(16)$  & $6.7(4.0)$  & $12(10)$  & $-3.827(90)$  & $0.69(10)$  & $14.5(1.8)$ \\
0.45 & 4.9/5  & $175.7(1.9)$  & $-53(18)$  & $-3.17(78)$  & $-0.08(25)$  & $10.1(4.3)$  & $8(11)$  & $-3.716(72)$  & $0.68(11)$  & $18.5(2.0)$ \\
0.50 & 4.6/5  & $175.7(1.9)$  & $-55(17)$  & $-2.6(1.4)$  & $-0.26(42)$  & $12.1(4.9)$  & $6(11)$  & $-3.438(26)$  & $0.74(13)$  & $22.4(2.5)$ \\

    \end{tabular}
\end{ruledtabular}
\end{table}
}

{
\squeezetable
\begin{table}[h]
\caption{\label{table:KSWfits}
Prior and posterior values of parameters of the KSW extrapolation in $\overline{\rm MS}$ scheme.  $M_0$ is given in GeV, $c_1$ in ${\rm GeV}^{-1}$, $C_0^I$ in ${\rm fm}^{2}$ and $C_2^I$ and $D_2^I$ in ${\rm fm}^{4}$ units.
$\chi^2_\mathrm{aug}$ is the standard $\chi^2$ plus contributions from the priors~\cite{Lepage:2001ym}.}
\begin{ruledtabular}

\begin{tabular}{c c c c c c c c c c c}
&$\chi^2_\mathrm{aug}/\mathrm{dof}$ &
$M_0$& $c_1$& $g_{\pi NN}$ & $C_0^{I=0}$  & $D_2^{I=0}$  & $C_2^{I=0}$ & $C_0^{I=1}$  & $D_2^{I=1}$  & $C_2^{I=1}$ \\
\hline
Priors  & -- & 0.89(10)& -1.1(1.0)&13.3 (1.3)  & 0(100)& 0(100)& 0(100)& 0(100)& 0(100)& 0(100)  \\
Posteriors& 24.7/15 
&0.8919(49) & -0.780(91)&11.2 (1.1)
&14.6(6.1) &52(47)& 52(34)
&13.3(5.6) &76(63)& 76(58) \\
\end{tabular}
\end{ruledtabular}
\end{table}
}

\clearpage
\bibliography{cited_refs}   

\end{document}